\documentclass[twocolumn,aps,pre,floatfix,superscriptaddress,longbibliography]{revtex4-1}
\pdfoutput=1 
\usepackage{amsmath,amssymb,eucal,graphicx,float,epstopdf,xparse}
\usepackage{mathtools,xfrac,esint}  
\usepackage{epsfig,subfigure}
\usepackage[utf8]{inputenc}

\usepackage[colorlinks=true, urlcolor=blue, anchorcolor=blue, citecolor=blue,filecolor=blue,linkcolor=blue,menucolor=blue]{hyperref}

\begin{document}
\title{Hiring Strategies}

\author{P. L. Krapivsky}
\affiliation{Department of Physics, Boston University, Boston, Massachusetts 02215, USA}
\affiliation{Santa Fe Institute, Santa Fe, New Mexico 87501, USA}

\begin{abstract} 
We investigate the hiring problem where a sequence of applicants is sequentially interviewed, and a decision on whether to hire an applicant is immediately made based on the applicant's score. For the maximal and average improvement strategies, the decision depends on the applicant's score and the scores of all employees, i.e., previous successful applicants. For local improvement strategies, an interviewing committee randomly chosen for each applicant makes the decision depending on the score of the applicant and the scores of the members of the committee. These idealized hiring strategies capture the challenges of decision-making under uncertainty. We probe the average score of the best employee, the probability of hiring all first $N$ applicants, the fraction of superior companies in which, throughout the evolution, every hired applicant has a score above expected, etc. 
\end{abstract}

\maketitle

\section{Introduction}
\label{sec:Intro}

Imagine a growing company continuously interviewing and hiring applicants. We analyze three hiring strategies. In every strategy considered here, each applicant's quality is determined by a single number scored by the applicant during the interview. We further assume that each applicant is immediately accepted or rejected. The hiring decision relies on the score of the applicant and the set of scores of all employees or the members of an interviewing committee. 

Using a short interview to ascertain the quality of an applicant, in the long run, is unrealistic. Judging an applicant by a single number is questionable as there are different facets of the business. One could mimic the effectiveness of an applicant with a vector with components reflecting the strength of each potential task. The dimension of such vectors is fuzzy as there may be a need for employees who excel in a single task, two tasks, etc. The acceptance criteria are then difficult to define. Still, highly idealized hiring strategies capture the challenges of decision-making under uncertainty and shed light on more realistic hiring strategies.  

Hiring problems belong to a large class of decision making and optimal stopping problems, see, e.g., \cite{Miller,Lindley,Dynkin,Robbins64,Robbins65,Gusein-Zade, Konheim,Robbins,Siegmund,Berezovsky,Knuth-2,Kawai,Groot,decision,KR-parking-1,KR-parking-2}. The classical example is the secretary problem \cite{Miller,Lindley,Dynkin,Robbins64,Freeman,Ferguson,Bruss00}. Applicants for a secretarial position are interviewed one-by-one, and each applicant is compared to previous (rejected) applicants. An applicant is either accepted (and the process stops) or rejected. The secretary problem admits numerous modifications \cite{Vanderbei,Samuels76,Rose,Bruss87, Kleinberg-secretary,postdoc,Morris,Prophet,Zeevi,Gnedin}. 

A single decision is made in secretary problems. Decisions are continuously made in parking problems \cite{KR-parking-1,KR-parking-2} where the departure of agents compensates for their inflow, so the outcome is a fluctuating but statistically stationary state. We consider the inherently growing stage when a company continuously hires applicants; we ignore the layoff, leaving the company and retiring. 

The first applicant is effectively the founder of the company. The following applicants are hired or rejected according to the strategy-dependent rules that only depend on the characteristics of the employees. We shall analyze three simple strategies:
\begin{enumerate}
\item {\bf Maximal Improvement Strategy (MIS)}. To be accepted, the applicant must get a higher score than any previously accepted applicant.
\item {\bf Average Improvement Strategy (AIS)}. To be accepted, the applicant must have the score above the average score of all the employees.
\item {\bf Local Improvement Strategy (LIS)}. A randomly chosen employee interviews each applicant. To be accepted, the applicant must score higher than the interviewer. More generally, a hiring committee of $c$ randomly chosen employees interviews each applicant. A successful applicant must score higher than every member of the committee.
\end{enumerate}
For the MIS and AIS, each hire improves the average score of the employees. In the case of the LIS, the new hire can result in a decrease in the average score. 

We always assume that the scores of the applicants are random quantities drawn independently from the same distribution. The strategies are deterministic, so the randomness of the scores is the only source of stochasticity for the MIS and AIS. In the case of the LIS, there is an additional source of stochasticity as a hiring committee for each applicant is randomly chosen.

The notion of the ``hiring problem" has been proposed by Broder et al. \cite{Broder} where the authors explored a few hiring strategies. Several strategies appeared in earlier work as selection rules. The MIS selection rule corresponds to records \cite{Records,Records-prob}. In the realm of records, the length of the sequence, i.e., the number of interviews, plays the crucial role. For the hiring via the MIS, the number of successful hires is important as we typically compare companies of equal size. The AIS selection rule goes back to \cite{Preater}. For rank-based strategies, see \cite{Krieger,Panholzer13,Panholzer14,Janson}. For more complicated modeling involving game-theoretical aspects, see \cite{Kleinberg}. 

We consider companies that continuously interview and hire applicants. The growth is stochastic, so to make fair testing between random outcomes we compare companies when their sizes are identical, say $n$. We specifically analyze the following characteristics:
\begin{enumerate}
\item The score $x_n$ of the last hired applicant, in particular, the average score $\langle x_n\rangle$.
\item The maximal score $m_n$, i.e., the score of the best employee.
\item If the best employee was hired at the $n-j_n$ successful interview, we want to know $j_n$. 
\item The fraction $P_n$ of superior companies of size $n$. Equivalently, $P_n$ is the probability that throughout the evolution every hired applicant has the score above expected: $x_j>\langle x_j\rangle$ for $j=1,\ldots,n$. 
\item The probability $F_N$ that the company has reached size $N$ by hiring all first $N$ applicants, that is, not a single applicant has been rejected. 
\end{enumerate}

The maximal improvement strategy (MIS) is the most tractable (Sec.~\ref{sec:MIS}). For this strategy, 
\begin{equation}
\label{MIS:S-N}
x_n = m_n, \quad j_n=0, \quad F_N=(N!)^{-1}
\end{equation}
immediately follow from the definition of the MIS. These results hold independently on the score distribution. The fractions $P_n$ of superior companies depend on the score distribution. For the uniform distribution 
\begin{equation}
\label{uniform}
\rho(x) = 
\begin{cases}
1  & 0<x<1\\
0  & x>1
\end{cases}
\end{equation}
$\langle x_j\rangle = 1-2^{-j}$ and the fractions of superior companies admit an integral representation   
\begin{equation}
\label{Pn:MIS}
P_n = n!\idotsint\displaylimits_{\max(1-2^{-j},x_{j-1})<x_j<1}dx_1\ldots dx_n
\end{equation}
with $x_0\equiv 0$. The normalization factor $n!$ can be understood combinatorially or geometrically from 
\begin{equation}
P_n = \frac{\idotsint\displaylimits_{\max(1-2^{-j},x_{j-1})<x_j<1}dx_1\ldots dx_n}{\idotsint\displaylimits_{0<x_1<\cdots<x_n<1}dx_1\ldots dx_n}
\end{equation}
and an observation that the integral in the denominator, the volume of the $n-$dimensional simplex, is $1/n!$.

The fractions $P_n$ are rational numbers. The exact calculations of integrals \eqref{Pn:MIS} up to $n=8$ led us to the conjectural general answer
\begin{equation}
\label{Dn:MIS}
P_n = \frac{D_n}{2^{n^2}}
\end{equation}
where $D_n$ is a number of acyclic digraphs with $n$ labeled vertices \cite{Robinson70,Robinson73,Stanley,Harary,Rodionov,Stanley-I}. (A digraph is a graph with at most one directed edge from $i$ to $j$ for all $1\leq i,j\leq n$; acyclic means that there are no cycles.) The same quantity $D_n$ counts the number of $n\times n$ (0,1)-matrices whose eigenvalues are positive real numbers \cite{McKay}; see \cite{Sloane} for other interpretations of $D_n$. Thus according to \eqref{Dn:MIS}, the fraction  of superior companies of size $n$ is equal to the fraction of $n\times n$ matrices with $(0,1)$ elements whose eigenvalues are positive real numbers. 

Other strategies are less understood. In Sec.~\ref{sec:AIS}, we analyze the average improvement strategy (AIS). For instance, we determine the probability 
\begin{equation}
\label{SN-prod:AIS}
F_{N} = \prod_{j=1}^{N-1}\left[1 -\left(\frac{j}{j+1}\right)^{N-j}\right]
\end{equation}
that the company has reached size $N$ by hiring all first $N$ applicants. 

In Sec.~\ref{sec:LIS}, we consider local improvement strategies LIS($c$). Even the simplest such strategy with a hiring committee consisting of a single employee, $c=1$, is very challenging, and we present exact calculations only for $n\leq 3$. The probabilities $F_N$, in contrast, appear simple for all $c$.
 
 In Sec.~\ref{sec:best}, we study the properties of the best employee, particularly the average score $\langle m_n\rangle$ and average age $\langle j_n\rangle$. These characteristics are easy to compute for the MIS but appear analytically intractable for other strategies. We employ heuristic arguments to probe the asymptotic behaviors of $\langle m_n\rangle$ and $\langle j_n\rangle$ for the AIS and LIS(1). 

We use the uniform distribution \eqref{uniform} if not stated otherwise. To illustrate the effect of the score distribution, we present a few results for another compact score distribution, the tent distribution $\rho(x)=2(1-x)$. Score distributions with non-compact support qualitatively change some behaviors, and in Appendix~\ref{ap:exp}, we repeat several calculations using the exponential distribution.

\section{Maximal Improvement Strategy}
\label{sec:MIS}

The maximal improvement strategy (MIS) postulates that applicants are interviewed one by one, and an applicant with a score exceeding the scores of all employees is immediately accepted; otherwise, an applicant is rejected. The scores are independently drawn from the same distribution. We require that the distribution does not contain delta functions so that the probability of a tie between the scores of two applicants is zero. The precise form of the score distribution is irrelevant, sometimes surprisingly, for some results. In derivations of these results, it is often convenient to rely on the uniform distribution \eqref{uniform} on the interval $(0,1)$, so we tacitly assume that the score distribution is uniform and explicitly state when we consider different score distributions. For instance, for the MIS, the probabilities $P_n$ depend on the score distribution, and in  Appendix~\ref{ap:MIS-exp}, we determine $P_n$ for the exponential score distribution. 

For the MIS, the scores $x_1,\ldots,x_n$ of the employees are linearly ordered:
\begin{equation}
\label{x:MIS}
0<x_1<\ldots<x_n<1
\end{equation}
The upper bound may be set to unity and interpreted as the score of the founder implying that the employees are less talented (the founder is not considered an employee). The uniform score distribution is the simplest compact distribution. The MIS rules make sense as long as the scores of applicants are compared to the scores of employees, not the founder. 

As the number of hired applicants increases, their quality approaches to the maximal excellency: $x_n\to 1$ as $n\to\infty$. To quantitatively probe the quality of the consecutive successful applicants, we note that the score $x_{j+1}$ is uniformly distributed on the interval $(x_j,1)$. Therefore
\begin{equation}
\label{av:MIS}
\langle x_{j+1}\rangle = \left \langle \frac{x_j+1}{2}\right\rangle = \frac{\langle x_{j}\rangle+1}{2}
\end{equation}
Denoting by $\xi_j$ the average gap between the score of the $j^\text{th}$ employee and the maximal excellency
\begin{equation}
\label{gap:def}
\xi_j = 1 -\langle x_{j}\rangle 
\end{equation}
we recast \eqref{av:MIS} into $\xi_{j+1} = \xi_j/2$, from which
\begin{equation}
\label{gap:MIS}
\xi_j = 2^{-j}
\end{equation}
Thus in the case of the MIS, the approach to the maximal excellency is exponential. 

For the MIS, the best employee is the last hired applicant, so $j_n=0$ and  $x_n = m_n$. The score of the last (and best) employee averaged over all realizations is 
\begin{equation}
\label{best:MIS}
\langle x_n\rangle = \langle m_n\rangle =1 - 2^{-n}
\end{equation}

The average score of the company with $n$ employees is 
\begin{equation}
\label{an:def}
a_n=\frac{1}{n}\sum_{j=1}^n x_j
\end{equation}
We also introduce the gap between the average score and the maximal excellency averaged over all realizations
\begin{equation}
\label{av-gap:def}
\mu_n = 1 -\langle a_n\rangle = \frac{1}{n}\sum_{j=1}^n \xi_j
\end{equation}
Using \eqref{gap:MIS} we obtain
\begin{equation}
\label{av-gap:MIS}
\mu_n =  \frac{1}{n}\left(1-\frac{1}{2^n}\right)
\end{equation}
implying the algebraic $n^{-1}$ approach to the maximal excellency.

The probability  $F_N$  that the first $N$ applicants are all hired is given by
\begin{equation}
\label{SN:MIS}
F_N=\idotsint\displaylimits_{0<x_1<\cdots<x_N<1}dx_1\ldots dx_N = \frac{1}{N!}
\end{equation}
in agreement with \eqref{MIS:S-N}. The above derivation relies on the fact that the integral in \eqref{SN:MIS} is the volume of the simplex $0<x_1<\ldots<x_N<1$. But the answer is universal, that is, independent of the distribution as long as the probability of ties is zero. Indeed, there are $N!$ different orderings of the scores of the first $N$ applicants, so $1/N!$ is indeed the probability that the ordering is ascending.

\begin{figure}[ht]
\begin{center}
\includegraphics[width=0.4\textwidth]{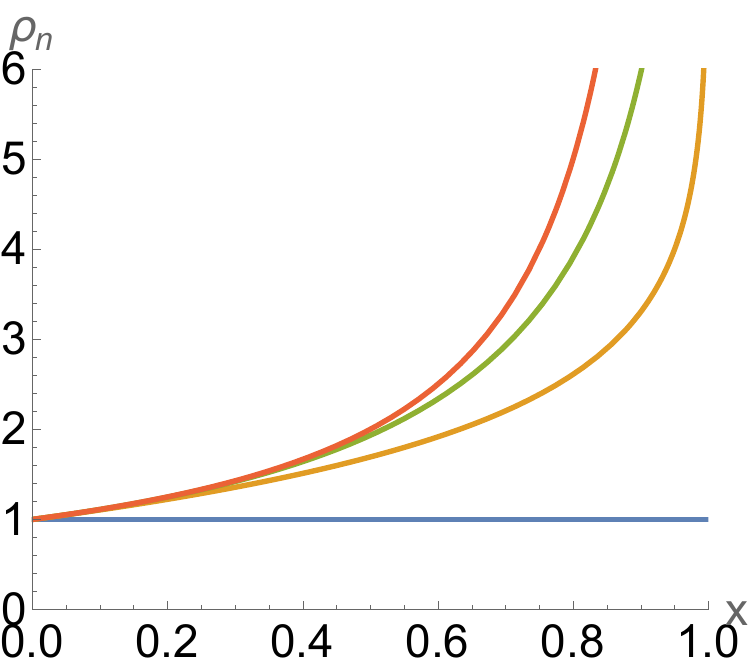}
\caption{The densities of scores $\rho_n(x)$ of the company with $n$ employees hired via the MIS for $n=1,2,3,\infty$ (bottom to top).}
\label{fig:r123}
  \end{center}
\end{figure}

The scores of employees satisfy \eqref{x:MIS}. Denote by $\rho_n(x)$ the density of scores. The normalization gives
\begin{equation}
\label{rho:norm}
\int_0^1 dx\,\rho_n(x) = n
\end{equation}
The density of scores depends on the `bare' distribution of scores, e.g., $\rho_1(x)=\rho(x)$. For our canonical uniform distribution \eqref{uniform}, we have
\begin{subequations}
\begin{equation}
\label{R-1}
\rho_1(x) = 1
\end{equation}
and
\begin{equation}
\label{R-2}
\rho_2(x) = \rho_1(x)+\int_0^x \frac{dx_1}{1-x_1} = 1 - \ln(1-x)
\end{equation}
When $n=3$, we similarly derive
\begin{eqnarray}
\rho_3(x) &=& \rho_2(x) + \int_0^x \frac{dx_1}{1-x_1}\int_{x_1}^x \frac{dx_2}{1-x_2} \nonumber\\
                &=& 1 - \ln(1-x) + \frac{[\ln(1-x)]^2}{2}
\end{eqnarray}
\end{subequations}
Generally
\begin{eqnarray}
\rho_{n+1}(x) = \rho_n(x) +  \int_{0<x_1<\ldots<x_n<x} \prod_{j=1}^n \frac{dx_j}{1-x_j}
\end{eqnarray}
from which
\begin{equation}
\rho_{n+1}(x) = \sum_{j=0}^{n} \frac{[-\ln(1-x)]^j}{j!}
\end{equation}
Thus $\rho_{n+1}$ diverges as $[-\ln(1-x)]^{n}$ when $x\to 1$, see Fig.~\ref{fig:r123}. The limiting distribution has a neat form
\begin{equation}
\label{R-inf}
\rho_\infty(x) = \frac{1}{1-x}
\end{equation}

In a {\em superior} company, every hired applicant has the score above the corresponding average score, $x_j >  \langle x_j\rangle$. Denote by $P_n$ the fraction of superior companies of size $n$. The scores of its employees must obey 
\begin{equation}
\label{def:superior}
x_j >  \langle x_j\rangle = 1 - 2^{-j}\,, \quad j=1,\ldots,n
\end{equation}
in the case of our canonical uniform distribution of scores. Relations \eqref{def:superior} together with \eqref{x:MIS} explain the integral representation \eqref{Pn:MIS}. The factorial factor in \eqref{Pn:MIS} accounts for normalization: We must divide the integral in \eqref{Pn:MIS} by $1/n!$ reflecting  the linear order \eqref{x:MIS}. Using Mathematica, we computed the integrals \eqref{Pn:MIS} for $n\leq 4$ and observed the announced result \eqref{Dn:MIS}. Further calculations confirmed \eqref{Dn:MIS} up to $n=8$. Equations  \eqref{Pn:MIS}--\eqref{Dn:MIS} imply an integral representation for $D_n$ which we have not found in published literature. The best known way to compute $D_n$ relies on a beautiful recurrence \cite{Robinson73,Harary} which can be recast into recurrence for the probabilities
\begin{equation}
\label{Pn:rec}
P_n = \sum_{k=1}^n (-1)^{k-1}\binom{n}{k}\, 2^{-nk}\, P_{n-k}
\end{equation}
with $P_0=1$. Reducing the integral representation \eqref{Pn:MIS} to the recurrence \eqref{Pn:rec} would constitute the proof of \eqref{Dn:MIS}. 

Plugging an asymptotic formula for $D_n$ established by Robinson \cite{Robinson70,Robinson73} and Stanley \cite{Stanley} into the conjectural exact formula \eqref{Dn:MIS} we obtain
\begin{equation}
\label{Pn:asymp}
P_n \simeq \frac{n!}{M p^n}\,\, 2^{-\frac{n(n+1)}{2}}
\end{equation}
with $p = 1.488\ldots$ and $M = 0.474\ldots$ for $n\gg1$. 

The MIS selection rule corresponds to records, viz., consecutive maxima, in a sequence of random variables. In the realm of records, all variables are treated on the same footing \cite{Records,Records-prob}, while according to the MIS only applicants with currently maximal score are employed. Superior probabilities appeared in the realm of records \cite{BK-13} suggesting our definition of $P_n$. 

The probabilities $P_n$ depend on the score distribution. In Appendix~\ref{ap:MIS-exp}, we probe the probabilities $P_n$ for the exponential score distribution.

\section{Average Improvement Strategy}
\label{sec:AIS}

The average improvement strategy (AIS) posits that an applicant with a score exceeding the average score of the employees is immediately accepted; otherwise, an applicant is rejected. The AIS goes back to the paper by Preater \cite{Preater} who used different terminology; the notion of the hiring problem was introduced later \cite{Broder}. The AIS and similar strategies were studied in Refs.~\cite{Broder,Krieger,Panholzer13,Panholzer14,Janson}. 

We begin with a few basic properties of the AIS that can be extracted from earlier work. We then establish the general formula for the fraction of superior companies and derive Eq.~\eqref{SN-prod:AIS} giving the probabilities $F_N$. We also briefly discuss the densities $\rho_n(x)$ and the limiting density $\rho_\infty(x)$. In contrast to the MIS where the densities $\rho_n(x)$ are smooth, in the case of the AIS, the density $\rho_n(x)$ has singularities at $x=1/k$ with $k=2,\ldots,n-1$. 

The definition of the AIS implies that the scores satisfy
\begin{equation}
x_{k+1}>a_k 
\end{equation}
for $k=1,\ldots,n-1$. Therefore, the score $x_{k+1}$ is uniformly distributed on the interval $(a_k, 1)$, and we have (cf. with Eq.~\eqref{av:MIS} for the MIS)
\begin{equation}
\label{av:AIS}
\langle x_{k+1}\rangle = \frac{\langle a_k\rangle+1}{2}
\end{equation}
Using $x_{k+1}=(k+1)a_{k+1}-k a_k$ following from the definition \eqref{an:def} we deduce 
\begin{equation}
\label{av-xa:AIS}
\langle x_{k+1}\rangle = (k+1)\langle a_{k+1}\rangle - k \langle a_k\rangle
\end{equation}
Combining \eqref{av:AIS} and \eqref{av-xa:AIS} we obtain
\begin{equation}
\label{m-av:AIS}
(k+1)\langle a_{k+1}\rangle = \left(k+\tfrac{1}{2}\right)\langle a_k\rangle + \tfrac{1}{2}
\end{equation}
Using \eqref{av-gap:def} we recast \eqref{m-av:AIS} into $(k+1) \mu_{k+1} = \left(k+\tfrac{1}{2}\right)\mu_k$. The solution to this recurrence satisfying $\mu_1=\frac{1}{2}$ reads 
\begin{equation}
\label{rec-sol:AIS}
\mu_n = \frac{\Gamma\left(n+\frac{1}{2}\right)}{\Gamma\left(\frac{1}{2}\right)\Gamma(n+1)}
\end{equation}
Using \eqref{gap:def} and \eqref{av-gap:def} we re-write \eqref{av:AIS} as $\xi_{k+1}=\mu_k/2$ which in conjunction with \eqref{rec-sol:AIS} lead to
\begin{equation}
\label{gap:AIS}
\xi_n = \frac{\Gamma\left(n-\frac{1}{2}\right)}{2\Gamma\left(\frac{1}{2}\right)\Gamma(n)}
\end{equation}
The approach to the maximal excellency is algebraic:
\begin{equation}
\label{gap-asymp:AIS}
\mu_n \simeq  \frac{1}{\sqrt{\pi}}\,n^{-\frac{1}{2}}\quad\text{as}\quad n\gg 1
\end{equation}

Let us briefly look at general compact score distributions. If $\rho(x_\text{max})>0$, the approach to the maximal excellency resembles the asymptotic behavior \eqref{gap-asymp:AIS}, viz. $\mu_n\sim n^{-1/2}$ for $n\gg 1$. If $\rho(x)\sim (x_\text{max}-x)^{a}$ as $x\uparrow x_\text{max}$, the approach to the maximal density is $\mu_n\sim n^{-\frac{1}{2+a}}$. For instance, for the tent distribution 
\begin{equation}
\label{tent}
\rho(x)=
\begin{cases}
2(1-x)  & 0<x<1\\
0          & x>1
\end{cases}
\end{equation}
one finds
\begin{equation}
\label{tent:AIS}
\mu_n = \frac{\Gamma\left(n+\frac{2}{3}\right)}{\Gamma\left(\frac{2}{3}\right)\Gamma(n+1)}  \simeq  \frac{1}{\Gamma\left(\frac{2}{3}\right)}\,n^{-\frac{1}{3}}
\end{equation}

We now return to our canonical uniform score distribution and note that the fraction of superior companies of size $n$ is given by the $n-$folded integral 
\begin{equation}
\label{Pn:AIS}
P_n = \frac{\idotsint\displaylimits_{\max(1-\xi_j,a_{j-1})<x_j<1}dx_1\ldots dx_n}{\idotsint\displaylimits_{a_{j-1}<x_j<1}dx_1\ldots dx_n}
\end{equation}
Here $a_0\equiv 0$ and $a_m=(x_1+\ldots+x_m)/m$ for $m>0$; the quantities $\xi_j=1-\langle x_j\rangle$ are determined by \eqref{gap:AIS}. The fractions $P_n$ are rational numbers. One can compute $P_1$ and $P_2$ by hand. For larger $n$, we used Mathematica. Table~\ref{Tab:Pn} collects $P_n$ for $n\leq 5$. We could not decipher any pattern and guess the general formula for $P_n$. The presence of huge prime factors, see the numerator of $P_5$, makes unlikely the existence of such a formula. 

\begin{table}[h!]
\centering
\renewcommand{\arraystretch}{1.6}{
\begin{tabular}{|c|c|c|c|c|c|c|}
\hline
$n$ & $1$ & $2$ & $3$ & $4$ &$5$  \\
\hline
$P_n$ & $\frac{1}{2}$ & $\frac{3}{2^4}$ & $\frac{3^2\cdot 7}{2^{10}}$ & $\frac{3^2\cdot 43\cdot 173}{2^{19}\cdot 7}$  & $\frac{83\cdot 2\,051\,182\,663}{2^{34}\cdot 3\cdot 5\cdot 7\cdot 19}$  \\
\hline
\end{tabular}
}
\caption{The fractions $P_n$ of superior companies of small size, $n\leq 5$, for the AIS. The general expression is \eqref{Pn:AIS}. }
\label{Tab:Pn}
\end{table}

For the AIS, the probabilities $F_N$ depend on $\rho(x)$. The only exceptions are the first two probabilities 
\begin{equation}
\label{F12}
\begin{split}
F_1 &=\int_0^\infty dx_1\,\rho(x_1)=1\\
F_2 &=\int_0^\infty dx_1\,\rho(x_1) \int_{x_1}^\infty dx_2\,\rho(x_2)=\frac{1}{2}
\end{split}
\end{equation}
Starting from 
\begin{equation*}
F_3 =\int_0^\infty dx_1\,\rho(x_1) \int_{x_1}^\infty dx_2\,\rho(x_2) \int_{\frac{x_1+x_2}{2}}^\infty dx_3\,\rho(x_3)
\end{equation*}
the probabilities depend on $\rho(x)$. For instance, $F_3 = \frac{1}{4}$ for the uniform distribution, $F_3 = \frac{17}{72}$ for the tent distribution, and $F_3 = \frac{2}{9}$ for the exponential distribution.

We now derive the announced general formula \eqref{SN-prod:AIS} for the uniform distribution. The first $N$ applicants are all hired with probability
\begin{equation}
\label{SN:AIS}
F_N=\int_0^1 dx_1 \int_{a_1}^1 dx_2 \ldots  \int_{a_{N-1}}^1 dx_N
\end{equation}
One easily computes
\begin{equation}
\label{12345}
F_1=1, ~~ F_2=\tfrac{1}{2},  ~~  F_3 = \tfrac{1}{4},  ~~  F_4 = \tfrac{35}{288},  ~~  F_5 = \tfrac{133}{2304}
\end{equation}
The probabilities \eqref{SN:AIS} for the uniform distribution exceed the probabilities $F_N=1/N!$ for the MIS for all $N\geq 3$. In Table~\ref{Tab:SN}, we collect $F_N$ for $6\leq N\leq 9$. 

\begin{table}[h!]
\centering
\renewcommand{\arraystretch}{1.6}{
\begin{tabular}{|c|c|c|c|c|}
\hline
$N$& $6$ & $7$ & $8$ &$9$\\
\hline
$F_N$& $\tfrac{14\,911}{552\,960}$ 
& $\tfrac{991\,067}{79\,626\,240}$  & $\tfrac{13\,058\,067\,737}{2\,293\,235\,712\,000}$ & $\tfrac{3\,014\,412\,193\,738\,231}{1\,165\,037\,125\,238\,784\,000}$\\
\hline
\end{tabular}
}
\caption{The probabilities $F_N$ for $N=6,7,8,9$ for the AIS with uniform score distribution. When $N\leq 5$, the probabilities are given by \eqref{12345}; the general expression is \eqref{SN-prod:AIS}. The probabilities decay exponentially with an algebraic pre-factor, Eq.~\eqref{SN-asymp:AIS}. }
\label{Tab:SN}
\end{table}

To deduce the general formula we begin by noticing that the first integral in \eqref{SN:AIS} is 
\begin{subequations}
\label{SN:A}
\begin{equation}
\label{SN:A1}
\int_{a_{N-1}}^1 dx_N = 1- a_{N-1}
\end{equation}
Next one computes
\begin{equation}
\label{SN:A2}
\begin{split}
&\int_{a_{N-2}}^1 dx_{N-1}\,(1- a_{N-1}) = A_2^{(N)} (1- a_{N-2})^2\\
&A_2^{(N)}  = \tfrac{N-1}{2}\left[1-\big(\tfrac{N-2}{N-1}\big)^2\right]
\end{split}
\end{equation}
Continuing one gets
\begin{equation}
\label{SN:A3}
\begin{split}
&\int_{a_{N-3}}^1 dx_{N-2}\,(1- a_{N-2})^2 = A_3^{(N)}  (1- a_{N-3})^3 \\
&A_3^{(N)}  = \tfrac{N-2}{3}\left[1-\big(\tfrac{N-3}{N-2}\big)^3\right]
\end{split}
\end{equation}
and then
\begin{equation}
\label{SN:A4}
\begin{split}
&\int_{a_{N-4}}^1 dx_{N-3}\,(1- a_{N-3})^3 = A_4^{(N)}  (1- a_{N-4})^4 \\
&A_4^{(N)}  = \tfrac{N-3}{4}\left[1-\big(\tfrac{N-4}{N-3}\big)^4\right]
\end{split}
\end{equation}
\end{subequations}
Thus $F_N$ is $A_2^{(N)}  A_3^{(N)} \ldots A_{N-2}^{(N)} $ times the final integral which happens to be $1/N$. From Eqs.~\eqref{SN:A} one recognizes the general expression for the amplitudes $A_k^{(N)} $, viz.
\begin{equation}
\label{SN-Ak}
A_k^{(N)} = \tfrac{N-k+1}{k}\left[1-\big(\tfrac{N-k}{N-k+1}\big)^k\right]
\end{equation}
Massaging $F_N=N^{-1}A_2^{(N)}  A_3^{(N)} \ldots A_{N-2}^{(N)}$ with $A_k^{(N)}$ given by \eqref{SN-Ak} we arrive at the announced result \eqref{SN-prod:AIS}. Taking the logarithm of \eqref{SN-prod:AIS} and replacing summation by integration we find that $F_N$ exhibits an exponential decay:
\begin{subequations}
\label{SN-asymp:AIS}
\begin{equation}
\lim_{N\to\infty}N^{-1} \ln F_N = -\Lambda 
\end{equation}
with
\begin{equation}
\Lambda = -\int_0^1 dy\,\ln\!\big(1-e^{1-1/y}\big) = 0.8433021075\ldots 
\end{equation}
Using the Euler-Maclaurin formula \cite{Knuth}, one additionally extracts an algebraic pre-factor:
\begin{equation}
\label{AIS:SN}
F_N \sim N^{-\frac{1}{2}} \, e^{-\Lambda N}
\end{equation}
\end{subequations}

Recall that for the MIS,  the probabilities $F_N$ are universal, i.e., independent on the score $\rho(x)$, and given by $F_N=1/N!$. We already demonstrated that for the AIS, the probability $F_3$ already depends on the score distribution. In Appendix~\ref{ap:AIS-exp}, we show that 
\begin{equation}
\label{SN:AIS-exp}
F_N = \frac{N!}{N^N}
\end{equation}
for the exponential score distribution. The decay is exponential with an algebraic pre-factor,
\begin{equation}
F_N \simeq e^{-N}\sqrt{2\pi N}\qquad\text{for}\quad N\gg 1,
\end{equation}
as in the case of the uniform score distribution.

\begin{figure}[ht]
\begin{center}
\includegraphics[width=0.44\textwidth]{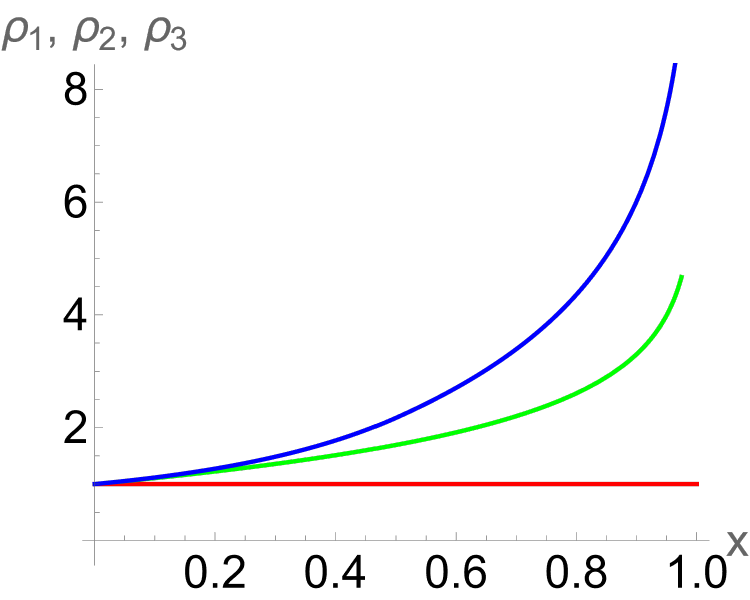}
\caption{The probability densities $\rho_1=1$, $\rho_2=1-\ln(1-x)$, see \eqref{R-2}, and $\rho_3(x)$ given by \eqref{R-3}. The second derivative of $\rho_3(x)$ is discontinuous at $x=\frac{1}{2}$ but this singularity is essentially invisible.}
\label{fig:r123-AIS}
  \end{center}
\end{figure}

The densities $\rho_1(x)$ and $\rho_2(x)$ are the same as before: \eqref{R-1} and \eqref{R-2}, respectively. It is still possible to find $\rho_3(x)$ using straightforward calculations. The rules of the AIS imply
\begin{equation*}
\rho_3(x) = \rho_2(x) + \int_0^x \frac{dy}{1-y}\int_y^{\text{min}(2x-y,1)} \frac{dz}{1-\frac{y+z}{2}} 
\end{equation*}
from which we find (Fig.~\ref{fig:r123-AIS})
\begin{equation}
\label{R-3}
\rho_3 =
\begin{cases}
1 - \ln(1-x) + [\ln(1-x)]^2         & x<\frac{1}{2}\\
1-(\ln 2)^2 -[1+\ln 4] \ln(1-x)   & x>\frac{1}{2}
\end{cases}
\end{equation}

Finding a compact formula for $\rho_n(x)$ applicable to all $n$ seems impossible. Indeed, $\rho_3(x)$ is already non-smooth at the middle point, $x=\frac{1}{2}$. Generally $\rho_{n+1}(x)$ has singular points $\frac{1}{n}\,, \frac{1}{n-1}\,, \ldots, \frac{1}{3}\,,\frac{1}{2}$. The score distribution $\rho_{n+1}(x)$ admits a neat general solution
\begin{equation}
\label{r-small-AIS}
\rho_{n+1}(x) = \sum_{j=0}^{n} [-\ln(1-x)]^j
\end{equation}
in the sub-interval $0<x<\frac{1}{n}$.

The limiting distribution $\rho_\infty(x)$ has singularities at $x=\frac{1}{k}$ with $k$ running over all natural numbers. Using \eqref{r-small-AIS} one guesses that the small $x$ expansion of $\rho_\infty(x)$ coincides with the small $x$ expansion of $[1+\ln(1-x)]^{-1}$, that is
\begin{equation}
\rho_\infty(x) = 1 + x + \frac{3x^2}{2} +  \frac{7x^3}{3} +  \frac{11x^4}{3} +  \frac{347 x^5}{60} + \ldots
\end{equation}
For the MIS, the limiting distribution \eqref{R-inf} diverges when $x\to 1$. The same is expected in the present case since $\int_0^1 dx\,\rho_\infty(x) = \infty$. To probe the divergence we use heuristic arguments. We begin with $\rho_1=\frac{1}{\mu_0}=1$. Then we can roughly estimate $\rho_2=\frac{1}{\mu_0}+\frac{1}{\mu_1}$ for $x>1-\mu_1$. Proceeding this way we get an estimate
\begin{equation}
\rho_\infty(1-\mu_n)= \sum_{j=0}^{n-1}\frac{1}{\mu_j}
\end{equation}
Using the asymptotic \eqref{gap-asymp:AIS} we get
\begin{equation}
\rho_\infty\!\left(1-\frac{1}{\sqrt{\pi n}}\right)= \frac{2}{3\pi}(\pi n)^{3/2}
\end{equation}
This leads to the following estimate for the score density
\begin{equation}
\label{R-n}
\rho_n(x) =
\begin{cases}
\frac{2}{3\pi}\,(1-x)^{-3}     & x<1-\frac{1}{\sqrt{\pi n}}\\
\frac{2}{3\pi}\,(\pi n)^{3/2} & x >1-\frac{1}{\sqrt{\pi n}}
\end{cases}
\end{equation}
The above heuristic derivation is meant to describe $\rho_n(x)$ for $n\gg 1$. The prediction \eqref{R-n} differs of course from the exact results for small $n$, see \eqref{R-1}, \eqref{R-2}, \eqref{R-3}. Interestingly, the approximate result \eqref{R-n} perfectly agrees with the exact normalization condition \eqref{rho:norm} and it also asymptotically agrees with another exact sum rule
\begin{equation}
\label{rho:gap}
\int_0^1 dx\,(1-x)\rho_n(x) = n\mu_n
\end{equation}
These arguments suggest that in the $1-x\to +0$ limit
\begin{equation}
\label{R-inf-AIS}
\rho_\infty(x) \simeq \frac{2}{3\pi}\,(1-x)^{-3} 
\end{equation}
The $(1-x)^{-3}$ asymptotic appears to be exact, while the correct amplitude may differ from $\frac{2}{3\pi}$.

\section{Local Improvement Strategies}
\label{sec:LIS}

The simplest local improvement strategy is the hiring procedure in which each applicant is interviewed by a single randomly selected employee. In contrast to the previous strategies, a hiring event may decrease the average score, which is impossible for the MIS and AIS. All three hiring strategies are identical up to $n=2$ when the scores obey
\begin{equation}
0< x_1 < x_2 < 1
\end{equation}
with $x_1$ uniformly distributed on the interval $(0,1)$ and $x_2$ uniformly distributed on the interval $(x_1,1)$. For $n\geq 3$, thre strategies are different. For the LIS with $n=3$, the score of the third employee averaged over all $x_3$, but with fixed $x_1$ and $x_2$, reads
\begin{eqnarray}
\label{3:12}
\langle x_3\rangle_{x_1,x_2} &=&\frac{1-x_1}{2-x_1-x_2}\,\frac{x_1+1}{2}+\frac{1-x_2}{2-x_1-x_2}\,\frac{x_2+1}{2}\nonumber\\
&=&\frac{1-(x_1^2+x_2^2)/2}{2-x_1-x_2}
\end{eqnarray}
Indeed, if the third employee is hired when the first [resp. second] employee was interviewing, the average score of the third employee is $\frac{x_1+1}{2}$ [resp. $\frac{x_2+1}{2}$]. The factor $\frac{1-x_1}{2-x_1-x_2}$ [resp. $\frac{1-x_2}{2-x_1-x_2}$] is the probability of succeeding when the first [resp. second] employee was interviewing. The average score $a_3=\frac{x_1+x_2+x_3}{3}$ of the company with three employees after averaging over $x_3$ becomes
\begin{equation}
\label{a3:12}
\langle a_3\rangle_{x_1,x_2} = \frac{x_1}{3}+\frac{x_2}{3}+\frac{1}{6}\,\frac{2-x_1^2-x_2^2}{2-x_1-x_2}
\end{equation}
Averaging \eqref{a3:12} over $x_2$ we obtain
\begin{eqnarray*}
\langle a_3\rangle_{x_1} = \frac{x_1}{3}+\frac{1+x_1}{6} + \frac{1}{6}\int_{x_1}^1 \frac{dx_2}{1-x_1}\,\frac{2-x_1^2-x_2^2}{2-x_1-x_2}
\end{eqnarray*}
Computing the integral we obtain
\begin{equation}
\langle a_3\rangle_{x_1} = \frac{1}{12}\left[7-4\ln 2 + x_1(5+4\ln 2) \right]
\end{equation}
which after the final averaging over $x_1$ yields
\begin{equation}
\langle a_3\rangle^\text{LIS} = \frac{19}{24} - \frac{1}{6}\,\ln 2
\end{equation}
Thus the gap $\mu_3 = 1 - \langle a_3\rangle$ is
\begin{equation}
\mu_3^\text{LIS} = \frac{5}{24} + \frac{1}{6}\,\ln 2 = 0.3238578634\ldots
\end{equation}
This gap exceeds the corresponding gap for the AIS
\begin{equation}
\mu_3^\text{AIS} = \frac{5}{16} = 0.3125
\end{equation}
which in turn exceeds the corresponding gap for the MIS
\begin{equation}
\mu_3^\text{MIS} = \frac{7}{24} = 0.29166666666\ldots
\end{equation}

An exact computation of $\mu_4^\text{LIS}$ is probably very cumbersome. We anticipate that the gap decays slower than for the  MIS and AIS, but still algebraically and with the same exponent as for the AIS:
\begin{equation}
\label{gap-asymp}
\mu_n \sim n^{-1/2} \quad \text{when}\quad n\gg 1
\end{equation}
The exponent is conjecturally the same since there is a lot of similarity between the AIS and LIS, more precisely the LIS(1), as we considered the LIS with a committee consisting of a single employee. 

The probabilities $F_1$ and $F_2$ are again given by \eqref{F12}. It proves convenient to recompute $F_1$ and $F_2$ via the cumulative distribution $R(x)=\int_x^\infty dy\,\rho(y)$:
\begin{equation}
\label{F12:R}
F_1  = \int_0^1 dR =1, \quad F_2 = \int_0^1 dR\,R=\frac{1}{2}
\end{equation}
The probability $F_3$ is given by 
\begin{eqnarray}
\label{F3}
F_3 &=& \frac{1}{2}\int_0^\infty dx_1\,\rho(x_1) \int_{x_1}^\infty dx_2\,\rho(x_2) \int_{x_1}^\infty dx_3\,\rho(x_3) \nonumber \\
&+& \frac{1}{2}\int_0^\infty dx_1\,\rho(x_1) \int_{x_1}^\infty dx_2\,\rho(x_2) \int_{x_2}^\infty dx_3\,\rho(x_3) \nonumber \\
&=& \frac{3}{2}\int_0^1 dR\,R^2=\frac{1}{4}
\end{eqnarray}
and also universal. The next probability is
\begin{eqnarray*}
6F_4 &=& \int_0 dx_1 \int_{x_1} dx_2 \int_{x_1} dx_3 \int_{x_1} dx_4 \,\rho_4(\mathbf{x}) \nonumber \\
         &+& \int_0 dx_1 \int_{x_1} dx_2 \int_{x_1} dx_3 \int_{x_2} dx_4 \,\rho_4(\mathbf{x}) \nonumber \\
         &+& \int_0 dx_1 \int_{x_1} dx_2 \int_{x_1} dx_3 \int_{x_3} dx_4 \,\rho_4(\mathbf{x}) \nonumber \\
         &+& \int_0 dx_1 \int_{x_1} dx_2 \int_{x_2} dx_3 \int_{x_1} dx_4 \,\rho_4(\mathbf{x}) \nonumber \\
         &+& \int_0 dx_1 \int_{x_1} dx_2 \int_{x_2} dx_3 \int_{x_2} dx_4 \,\rho_4(\mathbf{x}) \nonumber \\
         &+& \int_0 dx_1 \int_{x_1} dx_2 \int_{x_2} dx_3 \int_{x_3} dx_4 \,\rho_4(\mathbf{x}) 
\end{eqnarray*}
where $\rho_4(\mathbf{x})\equiv \rho(x_1)\rho(x_2)\rho(x_3)\rho(x_4)$ and we explicitly marked only the (non-trivial) lower limit of integration.  Using the cumulative distribution we re-write $F_4$ as  
\begin{equation}
\label{F4}
F_4 = \frac{1}{2}\int_0^1 dR\,R^3=\frac{1}{8}
\end{equation}
Thus, $F_4$ is also universal. All probabilities $F_N$ are universal and given by remarkably simple formula
\begin{equation}
\label{SN:LIS}
F_N = \frac{N}{2^{N-1}}\int_0^1 dR\,R^{N-1} = \frac{1}{2^{N-1}}
\end{equation}

To prove \eqref{SN:LIS}, consider a modified mLIS(1) strategy assuming the presence of the employee with score $x_0=0$. The committee with this employee is legitimate, but we account only employees with positive score. One can check that the probabilities $\Phi_N$ to hire all first $N$ applicants by the MLIS(1) are given by the same integral formulas as $F_{N+1}$ times $N+1$, that is, 
\begin{equation}
\label{Phi:LIS}
\Phi_N=(N+1)F_N=\frac{N+1}{2^N}
\end{equation}

We verified the predictions \eqref{SN:LIS} and \eqref{Phi:LIS} for small $N$. We now assume the validity of \eqref{SN:LIS} and \eqref{Phi:LIS}. After choosing $x_1$, the following $x_2, \ldots, x_{N+1}$ are chosen according to mLIS(1) with $x_1$ playing the role of $x_0$. The difference is that $x_2, \ldots, x_{N+1}$ belong to $(x_1,1)$, so the probability density of the success is $\Phi_N[R(x_1)]^N$. Hence
\begin{equation}
F_{N+1} = \int_0^1 dR\,\Phi_N R^N
\end{equation}
and using \eqref{Phi:LIS} we obtain $F_{N+1} = 2^{-N}$ thereby completing the proof by induction. 

Denote by LIS($c$) the strategy when a randomly chosen committee of $c$ employees interviews each new applicant. To be hired, the applicant should score higher than each committee member. If size of the company is $\leq c$, all employees participate in every hiring decision, and the LIS($c$) is identical to the MIS. 

Consider now the probabilities $F_N$. The first two probabilities $F_1$ and $F_2$ are again given by \eqref{F12}. The following probabilities are also universal. For the LIS(2)
\begin{subequations}
\begin{align}
\label{F3:2}
F_3  &= \int_0 dx_1 \int_{x_1} dx_2 \int_{x_2} dx_3\,\rho(x_1)\rho(x_2)\rho(x_3)\nonumber \\
        &= \frac{1}{2}\int_0^1 dR\,R^2 =\frac{1}{6}
\end{align}
and the next probability is         
\begin{align}
\label{F4:2}
F_4  &= \frac{1}{3}\int_0 dx_1 \int_{x_1} dx_2 \int_{x_2} dx_3 \int_{x_2} dx_4 \,\rho_4(\mathbf{x}) \nonumber \\
         &+ \frac{2}{3}\int_0 dx_1 \int_{x_1} dx_2 \int_{x_2} dx_3 \int_{x_3} dx_4 \,\rho_4(\mathbf{x}) \nonumber \\
         &=\frac{1}{3}\int_0^1 dR\,R^3\left[\frac{1}{3}+\frac{2}{6}\right]=\frac{1}{18}
\end{align}
\end{subequations}
The values $F_2=\frac{1}{2}$, $F_3=\frac{1}{6}$, and  $F_4=\frac{1}{18}$ suggest
\begin{equation}
\label{SN:LIS-2}
F_N = \frac{1}{2\cdot 3^{N-2}}
\end{equation}
for the LIS(2) when $N\geq 2$.

Generally for the LIS($c$)
\begin{equation}
\label{SN:LIS-c}
F_N = 
\begin{cases}
\frac{1}{N!}                                &  N\leq c\\
\frac{1}{c!\cdot (c+1)^{N-c}}     &  N\geq c
\end{cases}
\end{equation}
This is easy to establish in the $N\leq c+1$ range relying again on the cumulative distribution:
\begin{equation*}
F_N = \int_0^1 dR\,\frac{R^{N-1}}{(N-1)!}=\frac{1}{N!}
\end{equation*}
One then proves \eqref{SN:LIS-c} for $N\geq c+2$ by induction following the same lines as in the proof for the LIS(1). 

\section{Best employee}
\label{sec:best}

Below, we consider companies with  $n$ employees hired according to some strategy. The employee with the highest score is the best, and we analyze the score $m_n$ and the age $j_n$ of the best employee. (If an employee hired at the $n-j_n$ successful interview is the best, we call $j_n$ the age of the best employee.)

The score $m_n$ and the age $j_n$ of the best employee fluctuate from realization to realization of the (random) hiring process. These quantities are related via 
\begin{equation}
m_n = \text{max}(x_1,\ldots,x_n) = x_{n-j_n}
\end{equation}
When $n=1$ we have $m_1=x_1$ and $j_1=0$. In the case of the MIS we have $m_n=x_n$ and $j_n=0$. For the AIS we have $0\leq j_n\leq n-2$ when $n\geq 2$. For the LIS($c$) 
\begin{equation}
0 \leq j_n \leq n-c-1
\end{equation}
if $n> c+1$, and $j_n=0$ when $n\leq c+1$. 

The chief features of the random quantities $m_n$ and $j_n$ are their average values. Let us look at the gap between the average score of the best employee and the maximal excellency,  $\nu_n=1-\langle m_n\rangle$, and the average age of the best employee $\langle j_n\rangle$. In the case of the MIS
\begin{equation}
\nu_n=2^{-n}, \quad \langle j_n\rangle = j_n = 0
\end{equation}
For other strategies, the exact behaviors are unknown. We anticipate that $\nu_n$ and $\langle j_n\rangle$ scale algebraically:
\begin{subequations}
\begin{align}
\label{nu}
& \nu_n \sim n^{-\beta} \\
\label{age}
& \langle j_n\rangle \sim n^\alpha
\end{align}
\end{subequations}
when $n\gg 1$. We now support these algebraic behaviors for the AIS relying on heurisic arguments.  

First, we argue in favor of \eqref{nu}. Imagine that a company of size $n$ has just performed successful hiring. The score of the new employer must fall into the range $1-a_n$; if it falls into the range $1-m_n$, this new employee becomes the best. Therefore
\begin{equation}
\label{nn}
1-m_{n+1}=
\begin{cases}
1-m_n &\text{prob} \quad 1-\frac{1-m_n}{1-a_n}\\
r(1-m_n) &\text{prob} \quad \frac{1-m_n}{1-a_n}
\end{cases}
\end{equation}
where $r$ is uniformly distributed on $[0,1]$. Averaging \eqref{nn} we find 
\begin{equation}
\label{nu-rec}
\nu_{n+1} = \nu_n-\frac{1}{2}\,R_n, \quad R_n =\left\langle \frac{(1-m_n)^2}{1-a_n}\right\rangle
\end{equation}
We now express $R_n$ through the average values
\begin{equation}
R_n \approx \frac{(1-\langle m_n\rangle)^2}{1-\langle a_n\rangle}=\frac{\nu_n^2}{\mu_n}
\end{equation}
Substituting this uncontrolled approximation into \eqref{nu-rec} and treating $n$ as a continuous variable we recast \eqref{nu-rec} into a differential equation $\frac{d\nu_n}{dn} = - \frac{\nu_n^2}{2\mu_n}=-\frac{\sqrt{\pi n}}{2}\,\nu_n^2$; in the last step, we used the asymptotic \eqref{gap-asymp:AIS}. Solving the differential equation we arrive at the 
\begin{equation}
\label{nu_n}
\nu_n \simeq Bn^{-3/2}\, \quad B =   \frac{3}{\sqrt{\pi}}
\end{equation}
confirming \eqref{nu} and fixing $\beta=\frac{3}{2}$.

\begin{figure}[ht]
\begin{center}
\includegraphics[width=0.44\textwidth]{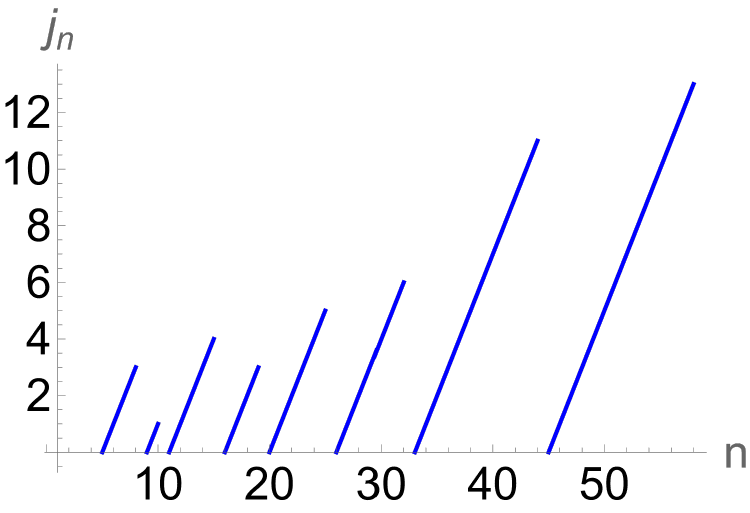}
\caption{Illustration of possible evolution of quantities $j_n$. In this example, $j_n=0$ for $1\leq n\leq 5$, implying that for $n\leq 5$ the last hired employee was the best. The first deterministic growth period occurs when $6\leq n\leq 8$.}
\label{fig:Jn}
  \end{center}
\end{figure}

To tackle $j_n$, we note that in a given realization, this random quantity increases by one at each step for a while and then drops to zero: The intervals of deterministic growth end in catastrophe, see Fig.~\ref{fig:Jn}. Let us try to estimate the duration of the interval between consecutive catastrophes. Suppose we start at some $n\gg 1$ such that $j_n=0$ and we want to estimate $\ell>n$ such that $j_\ell=0$ and $j_k=k-n$ when $n\leq k<\ell$. The probability that $j_k=k-n$ is estimated by a product 
\begin{equation}
\label{Pnk}
\Pi_n(k) = \prod_{j=n+1}^k \left(1- \frac{\nu_n}{\mu_j}\right)\simeq \exp\!\left[2-2\left(\frac{k}{n}\right)^{2/3}\right]
\end{equation}
The final result has been established using the asymptotic behaviors \eqref{gap-asymp:AIS} and \eqref{nu_n}. The average duration is then 
\begin{equation}
\label{age_n}
\begin{split}
&\int_n^\infty dk\,(k-n)\left(-\frac{d\Pi_n}{dk}\right) = \int_n^\infty dk\,\Pi_n(k)=Cn \\
&C =e^2\int_1^\infty dx\,e^{-2x^{2/3}} = 0.296\,788\ldots
\end{split}
\end{equation}
confirming \eqref{age} and fixing $\alpha=1$. 

The above arguments are heuristic, e.g., in Eq.~\eqref{Pnk}, we use the terms $1- \frac{\nu_n}{\mu_j}$ with evolving average gap $\mu_j$ and fixed gap $\nu_n$ between the average score of the best employee and the maximal excellency. The predictions $\beta=\frac{3}{2}$ and $\alpha=1$ for the exponents are probably correct, while the amplitudes $B$ and $C$ appearing in \eqref{nu_n} and \eqref{age_n} are less trustworthy. In addition to the average quantities $\nu_n=1-\langle m_n\rangle$ and $\langle j_n\rangle$, one would like to probe fluctuations, e.g., to understand the asymptotic behavior of the variance $\langle j_n^2\rangle_c = \langle j_n^2\rangle-\langle j_n\rangle^2$ of the age. 

\bigskip\noindent
I am grateful to Aaron Schweiger for useful suggestions.

\appendix
\section{Exponential score distribution}
\label{ap:exp}

Suppose the scores of applicants are independent identically distributed random variables drawn from the exponential distribution 
\begin{equation}
\label{exp}
\rho(x) = e^{-x}
\end{equation}
on the half-line $x\geq 0$. The shape of the distribution affects the outcome of the hiring process, while some details do not matter. For instance, for the two-parameter family of exponential distributions $\rho=\lambda e^{-\lambda(x-a)}$ on the half-line $x\geq a$, the results are the same as for \eqref{exp}. 
We now compute a few characteristics for the MIS and AIS with exponential score distribution.

\subsection{Maximal Improvement Strategy}
\label{ap:MIS-exp}
 
For the MIS with exponential score distribution \eqref{exp}, one finds
\begin{equation}
\label{best:MIS-exp}
\langle x_n\rangle = \langle m_n\rangle =n
\end{equation}

The fractions of superior companies admit an integral representation
\begin{equation}
\label{Pn:MIS-exp}
P_n = n!\idotsint\displaylimits_{\max(j,x_{j-1})<x_j<\infty}\prod_{j=1}^n e^{-x_j}dx_j
\end{equation}
with $x_0\equiv 0$. The probabilities can be expressed in a form somewhat resembling \eqref{Dn:MIS}, viz. 
\begin{equation}
\label{Dn:MIS-exp}
P_n = \frac{\mathcal{D}_n}{e^{n^2}}
\end{equation}
where $\mathcal{D}_n$ are polynomials of $e$ of degree $\frac{n(n-1)}{2}$ with integer coefficients. For instance
\begin{equation*}
\begin{split}
\mathcal{D}_1 & = 1 \\
\mathcal{D}_2 & = 2 e - 1 \\
\mathcal{D}_3 & = 6 e^3 - 6 e^2 + 1 \\
\mathcal{D}_4 & = 24 e^6 - 36 e^5 + 6 e^4 + 8 e^3 -1 \\
\mathcal{D}_5 & = 120 e^{10}  - 240 e^9 + 90 e^8 + 60 e^7 - 20 e^6 - 10 e^4 + 1  \\
\mathcal{D}_6 & = 720 e^{15}  - 1800 e^{14}  + 1080 e^{13} + 390 e^{12} - 360 e^{11} \\
                         &   -70 e^9 + 30 e^8 + 12 e^5 -  1 \\
\mathcal{D}_7 & =  5040 e^{21} -  15120 e^{20} +1260 e^{19}   + 1680 e^{18} \\
                       & -5040 e^{17} + 630 e^{16}  - 420 e^{15}  + 630 e^{14} \\
                       & - 70 e^{12} + 126 e^{11} - 42 e^{10} - 14 e^6 + 1 
\end{split}
\end{equation*}
etc. Three terms of the highest degrees appear to be
\begin{subequations}
\begin{equation}
\label{D-high}
\mathcal{D}_n = n!\,e^\frac{n(n-1)}{2}\big[1-\tfrac{n-1}{2e} + \tfrac{(n-2)(n-3)}{8e^2}+\ldots\big]
\end{equation}
with  third term contributing when $n\geq 4$. Two terms of the lowest degrees appear to be
\begin{equation}
\label{D-low}
\mathcal{D}_n = (-1)^{n-1} + 2n (-1)^n e^{n-1} +\ldots
\end{equation}
\end{subequations}
for $n\geq 3$.

Extracting  the asymptotic behavior of $P_n$ from the integral representation \eqref{Pn:MIS-exp} is an interesting challenge. The inequalities $x_j>j$ suggest the change of variables
\begin{equation}
\label{xy-exp}
x_j=j+y_j, \qquad j=1,\ldots,n
\end{equation}
turning \eqref{Pn:MIS-exp} into
\begin{equation}
\label{Pn:exp-y}
P_n = n!\, e^{-\frac{n(n+1)}{2}}\prod_{j=1}^n  \int_{(y_{j-1}-1)_+}^\infty e^{-y_j}dy_j
\end{equation}
where
\begin{equation}
z_+ = 
\begin{cases}
z  & z\geq 0\\
0  & z <0
\end{cases}
\end{equation}
Since $z_+\geq 0$, replacing the lower limits in all integrals in \eqref{Pn:exp-y} by zeros we obtain an upper bound
\begin{equation}
\label{Pn:exp-bound}
P_n \leq  n!\, e^{-\frac{n(n+1)}{2}}
\end{equation}
Comparing this bound with \eqref{Pn:asymp} hints that the multiple integral in \eqref{Pn:exp-y} has an asymptotic behavior 
\begin{equation}
\prod_{j=1}^n  \int_{(y_{j-1}-1)_+}^\infty e^{-y_j}dy_j \simeq \frac{1}{M p^n}
\end{equation}
Here we use the same notation as in \eqref{Pn:asymp}, but the values of $p$ and $M$ are different and we only know that  $p > 1$ and $M>0$. If true, the asymptotic behavior 
\begin{equation}
\label{Pn:asymp-exp}
P_n \simeq \frac{n!}{M p^n}\,\, e^{-\frac{n(n+1)}{2}}
\end{equation}
resembles the asymptotic \eqref{Pn:asymp} in the case of the uniform score distribution.

\subsection{Average Improvement Strategy}
\label{ap:AIS-exp}

For the AIS with exponential score distribution \eqref{exp}, the average score of the $(k+1)^\text{st}$ employee is 
\begin{equation}
\label{av:AIS-exp}
\langle x_{k+1}\rangle =  \langle a_k\rangle+1
\end{equation}
Combining \eqref{av:AIS-exp} and \eqref{av-xa:AIS} we obtain
\begin{equation}
\label{ak:AIS-exp}
\langle a_{k+1}\rangle - \langle a_k\rangle = \frac{1}{k+1}
\end{equation}
from which 
\begin{equation}
\label{ax:AIS-exp}
 \langle a_n\rangle = H_n, \qquad  \langle x_n\rangle = 1+H_{n-1}
\end{equation}
The harmonic numbers $H_n \equiv \sum_{1\leq k\leq n} k^{-1}$ grow logarithmically 
\begin{equation}
H_n = \ln n + \gamma  + \frac{1}{2n} -  \frac{1}{12n^2} + \ldots 
\end{equation}
where $\gamma=0.577\,215\ldots$ is the Euler constant \cite{Knuth}.

The probability that all first $N$ applicants are accepted admits an integral representation 
\begin{equation}
\label{FN:exp}
F_N=\prod_{j=1}^N \int_{a_{j-1}}^\infty dx_j\,e^{-x_j}
\end{equation}
with $a_0=0$ and $a_j=(x_1+\ldots+x_j)/j$ for $j>0$. Using $\int_{a_{N-1}}^\infty dx_N\,e^{-x_N}=e^{-a_{N-1}}$, we rewrite \eqref{FN:exp} as
\begin{equation}
\label{FNN:exp}
F_N=\prod_{j=1}^{N-1} \int_{a_{j-1}}^\infty dx_j\,e^{-Nx_j/(N-1)}
\end{equation}
from which
\begin{equation}
\label{FNN}
F_N=\left(\frac{N}{N-1}\right)^{N-1} F_{N-1}
\end{equation}
Starting from $F_1=1$ we deduce \eqref{SN:AIS-exp} from the recurrence \eqref{FNN}. 

The fractions of superior companies for $n\leq 2$ are
\begin{equation}
\label{P12}
\begin{split}
&P_1 = \frac{\int_1^\infty dx_1\,e^{-x_1}}{\int_0^\infty dx_1\,e^{-x_1}} = \frac{1}{e} \\
&P_2 = \frac{\int_1^\infty dx_1\int_{\max(x_1,2)}^\infty dx_2\,e^{-x_1-x_2}}{\int_0^\infty dx_1\int_{x_1}^\infty dx_2\,e^{-x_1-x_2}}=\frac{2e-1}{e^4}
\end{split}
\end{equation}
Generally, the fractions of superior companies admit an integral representation
\begin{equation}
\label{Pn:AIS-exp}
P_n = \frac{\idotsint\displaylimits_{\max(1+H_{j-1},a_{j-1})<x_j}\prod_{j=1}^n e^{-x_j}dx_j}{\idotsint\displaylimits_{a_{j-1}<x_j}\prod_{j=1}^n e^{-x_j}dx_j}
\end{equation}
with $a_0\equiv 0$. The denominator in \eqref{Pn:AIS-exp} is $F_n=n!/n^n$, so we can re-write \eqref{Pn:AIS-exp} as
\begin{equation}
\label{Pn:AIS-exp-factorial}
P_n = \frac{n^n}{n!}\idotsint\displaylimits_{\max(1+H_{j-1},a_{j-1})<x_j}\prod_{j=1}^n e^{-x_j}dx_j
\end{equation}
Using this integral representation one finds
\begin{equation}
\label{P3:AIS}
\begin{split}
P_3 &=\frac{18e^2 - 9e-14}{4e^{15/2}}=0.013\,071\,937\,993\ldots\\
P_4&=\frac{288e^3 -144e^2 - 224e-397}{27e^{34/3}}
\end{split}
\end{equation}
The results quickly become cumbersome, e.g.,
\begin{equation*}
P_5=\tfrac{540000 e^4 -270000 e^3 -420000e^2 - 744375 e-1448239}{20736 e^{185/12}}
\end{equation*}

\subsection{Local Improvement Strategies}
\label{ap:LIS}

For the LIS(1) with exponential score distribution, the average quantities $\langle a_n\rangle$ and $\langle x_n\rangle$ are given by \eqref{ax:AIS-exp}, i.e., the same as for the AIS. We also know $F_N$, see \eqref{SN:LIS}, as these probabilities do not depend on the score distribution. 

The integral representation for the fractions of superior companies quickly becomes cumbersome for large $n$. The first two fractions are given by \eqref{P12}. The third fraction is $P_3=N_3/D_3$ with
\begin{equation*}
\begin{split}
N_3 &= \int_1^\infty dx_1\int_{\max(x_1,2)}^\infty dx_2\int_{\max(x_1,5/2)}^\infty dx_3\,e^{-x_1-x_2-x_3}\\
        &+\int_1^\infty dx_1\int_{\max(x_1,2)}^\infty dx_2\int_{\max(x_2,5/2)}^\infty dx_3\,e^{-x_1-x_2-x_3}\\
D_3 &=\int_0^\infty dx_1\int_{x_1}^\infty dx_2\int_{x_1}^\infty dx_3\,e^{-x_1-x_2-x_3}\\
        &+\int_0^\infty dx_1\int_{x_1}^\infty dx_2\int_{x_2}^\infty dx_3\,e^{-x_1-x_2-x_3}
\end{split}
\end{equation*}
Computing $N_3$ and $D_3$ we find
\begin{equation}
\label{P3:LIS}
P_3=\frac{4e - e^{1/2}-2}{e^{13/2}}=0.010\,861\,455\ldots
\end{equation}
which is smaller than $P_3$ for the AIS, cf. \eqref{P3:AIS}. 

For the LIS(2) with exponential score distribution
\begin{equation}
\label{xn:n}
\langle x_n\rangle = n
\end{equation}
for $n\leq 3$. The computation of $\langle x_n\rangle$ for $n\geq 4$ is a bit more involved. Using the rules of the LIS(2), one gets
\begin{equation}
\label{xn:rec-long}
\langle x_{n+1}\rangle = 1 + \frac{1}{\binom{n}{2}}\sum_{j=1}^n (j-1)\langle x_j\rangle
\end{equation}
which we recast to the recurrence $\langle x_{n+1}\rangle = \langle x_n \rangle + \frac{2}{n}$ leading to
\begin{equation}
\label{xn:H}
\langle x_n\rangle = 2H_{n-1}
\end{equation}
for $n\geq 2$.

The fractions $P_1$ and $P_2$ are given by \eqref{P12}. The of the superior companies of size three
\begin{eqnarray}
\label{P3:LIS-2}
P_3 &=& \frac{\int_1^\infty dx_1\int_{\max(x_1,2)}^\infty dx_2\int_{\max(x_2,3)}^\infty dx_3\,e^{-x_1-x_2-x_3}}{\int_0^\infty dx_1\int_{x_1}^\infty dx_2\int_{x_2}^\infty dx_3\,e^{-x_1-x_2-x_3}} \nonumber\\
&=&\frac{6e^3 - 6e^{2}+1}{e^9}=0.009\,524\,631\ldots
\end{eqnarray}
is a little smaller than $P_3$ for the LIS(1), cf. \eqref{P3:LIS}. 

Generally for the LIS(c) with exponential score distribution, the average scores are easy to compute in the $n\leq c+1$ range where they are given by \eqref{xn:n}. The fractions of the superior companies are 
\begin{equation}
\label{Pn:LIS}
P_n = n!\idotsint\displaylimits_{\max(x_{j-1},j)<x_j}\prod_{j=1}^n e^{-x_j}dx_j
\end{equation}
when $n\leq c+1$. 

The fractions $P_1$ and $P_2$ are given by \eqref{P12} for all $c\geq 1$. The fraction $P_3$ is given by \eqref{P3:LIS-2} for all $c\geq 2$; the fraction $P_3$ is different, Eq.~\eqref{P3:LIS}, when $c=1$. The fraction of the superior companies of size four is
\begin{equation}
P_4 = \frac{24e^6-36e^5+6e^4 +8e^{3}-1}{e^{16}}
\end{equation}
for LIS($c$) with $c\geq 3$. The fraction of the superior companies of size five is
\begin{equation*}
P_5 = \tfrac{120 e^{10}-240 e^9 + 90 e^8 + 60 e^7 -20e^6-10 e^4 + 1}{e^{25}}
\end{equation*}
when $c\geq 4$, and the  fraction of the superior companies of size six is
\begin{equation*}
P_6\! =\!\tfrac{720 e^{15}  -1800 e^{14} + 1080 e^{13}+ 390 e^{12} - 360 e^{11}  - 70 e^9 + 30e^8 + 12 e^5 - 1}{e^{36}}
\end{equation*}
when $c\geq 5$.

\bibliography{references-hire}

\end{document}